S.M. Alkhimova*, S.V. Sliusar

Igor Sikorsky Kyiv Polytechnic Institute, Kyiv, Ukraine

*corresponding author: asnarta@gmail.com

# ANALYSIS OF EFFECTIVENESS OF THRESHOLDING IN PERFUSION ROI DETECTION ON T2-WEIGHTED MR IMAGES WITH ABNORMAL BRAIN ANATOMY

**Background.** The brain perfusion ROI detection being a preliminary step, designed to exclude non-brain tissues from analyzed DSC perfusion MR images. Its accuracy is considered as the key factor for delivering correct results of perfusion data analysis. Despite the large variety of algorithms developed on brain tissues segmentation, there is no one that works reliably and robustly on T2-weighted MR images of a human head with abnormal brain anatomy. Therefore, thresholding method is still the state-of-the-art technique that is widely used as a way of managing pixels involved in brain perfusion ROI in modern software applications for perfusion data analysis.
**Objective.** This paper presents the analysis of effectiveness of thresholding techniques in brain perfusion ROI detection on T2-weighted MR images of a human head with abnormal brain anatomy.
**Methods.** Four threshold-based algorithms implementation are considered: according to Otsu method as global thresholding, according to Niblack method as local thresholding, thresholding in approximate anatomical brain location, and brute force thresholding. The result of all algorithms is images with pixels' values changed to zero for background regions (air pixels and pixels that represent non-brain tissues) and original values for foreground regions (brain perfusion ROIs). The analysis is done using comparison of qualitative perfusion maps produced from thresholded images and from the reference ones (manual brain tissues delineation by experienced radiologists). The same DSC perfusion MR datasets of a human head with abnormal brain anatomy from 12 patients with cerebrovascular disease are used for comparison.
**Results.** Pearson correlation analysis showed strong positive ($r$ was ranged from 0.7123 to 0.8518, $p < 0.01$) and weak positive ($r < 0.35$, $p < 0.01$) relationship in case of conducted experiments with CBF, CBV, MTT and Tmax perfusion maps, respectively. Linear regression analysis showed at level of 95 % confidence interval that perfusion maps produced from thresholded images were subject to scale and offset errors in all conducted experiments.
**Conclusions.** The experimental results showed that widely used thresholding methods are an ineffective way of managing pixels involved in brain perfusion ROI. Thresholding as brain segmentation tool can lead to poor placement of perfusion ROI and, as a result, produced perfusion maps will be subject to artifacts and can cause falsely high or falsely low perfusion parameter assessment.

### Introduction

DSC (Dynamic Susceptibility Contrast) perfusion MR (Magnetic Resonance) imaging plays a significant role in diagnostic and management of cerebrovascular and intracranial oncological diseases [1–3]. During the DSC MR exam a scanner provides rapid acquisition of contrast-based image sequences to measure the first pass of a bolus as it circulates through the brain vasculature. The susceptibility of the contrast agent causes a decrease in signal intensity on the T2-weighted MR images. That kind of decrease in signal intensity is further converted into time-concentration curves from which perfusion data analysis can be performed on pixel-by-pixel basis. Result of DSC perfusion data analysis is quantitative values of hemodynamic parameters and perfusion maps which are their visual interpretation.

Nowadays, accurate detection of brain perfusion ROI is considered to be more relevant for delivering correct results of perfusion data analysis [4–6]. It can be explained by the fact that involving of non-target pixels data in perfusion analysis leads to the presence of numerous artifacts on perfusion maps and can cause falsely high or falsely low results of perfusion parameters assessment.

Accurate detection of brain perfusion ROI is a task of applying brain tissues segmentation algorithm that can provide proper results on DSC perfusion MR images with abnormal brain anatomy. Certainly, the manual segmentation is able to give accurate results not only on images with healthy tissues and organs, but with pathological ones too. Unfortunately, manual segmentation of brain tissues through all DSC exam images is labor intensive and extremely time-consuming task. In most clinical cases, its performance requires a specialist with sufficient knowledge and practical experience to detect brain anatomical structures and its lesions on T2-weighted MR images. Therefore, automatic algorithms for segmentation is generally preferable.

Despite the importance, the development of appropriate and effective automatic, or at least semi-automatic, brain segmentation algorithm on DSC perfusion MR images is still required for clinical use. The explanation for this is actually quite straightforward and can be substantiated by the following items.

Considering the fact that DSC perfusion data is T2-weighted images, such data is more complicated for automated brain segmentation than T1-weighted images. It can be explained with fatty tissues presence between brain and skull. Thus, a lot of automatic algorithms for brain segmentation are focused on T1-weighted MR images.

There are automatic algorithms that provide brain segmentation and simultaneously applicable to brain tissues detection on T1-weighted and T2-weighted MRI images [7, 8]. Such algorithms require a lot of parameters to be properly estimated for each, or at least initial, image processing. As a result, not accurate brain tissues ROIs are detected if the estimation and initialization are not done properly or in case of processing of low-contrast and low-resolution images, like DSC perfusion ones [9].

Any of the methods that are known in the art as useful for brain segmentation on T2-weighted MR images are not applicable to clinical use. It can be explained by required modifications in exam protocols (in case of utilizing for segmentation process the benefits from the specific image acquisition technique [10], or from the high-resolution pairs of T1-weighted and T2-weighted images [11]), or by partially eliminated information about studied objects (in case of parameterization of the T2-weighted image intensity onto a standardized T1-weighted intensity [12]).

Clinical images in most cases have visualization of abnormal brain anatomy. So, high amount of lesions types and their challenging shape or appearance in the brain are the reason of fails of automatic algorithm commonly applied for segmentation purposes in the field of computer software related to different areas of medical image processing [13—15]. In case of applying intensity based segmentation, like thresholding or clustering, incorrect results are caused by overlapping pixel intensities in lesion regions and regions which are targeted to be excluded from the image. In case of pattern recognition, at the present moment, there is a lack of pre-segmented templates and training samples for different shape, density, and location of the lesion for such algorithms applying on images with abnormal brain anatomy.

According to all the reasons mentioned above, thresholding method is still the state-of-the-art technique that is widely used as a way of managing pixels involved in brain perfusion ROI. For the most part, software applications for perfusion data analysis are oriented on automatic way of threshold value selection. However, on practice it is very common to have further manual turning of the threshold value to provide more accurate brain perfusion ROI detection. The basic principle of thresholding method is to divide pixels into two classes and thus differentiate the brain perfusion ROI from background. This principle is useful to implement intuitive user controls for threshold value selection, but, as was mentioned above, overlapping of pixel intensities in lesion regions and background regions can lead to incorrect brain perfusion ROI detection.

**Problem Statement**

The purpose of this study is to provide analysis of effectiveness of thresholding techniques in brain perfusion ROI detection on T2-weighted MR images of a human head with abnormal brain anatomy. This study focuses only on the threshold-based algorithms of low-level intensity pixels extraction that are widely used for medical image processing or specifically developed for brain segmentation on T2-weighted MR images.

The rest of the paper is organized as follows. The section Material and Methods presents the background for the experiments in the form of a description of threshold-based algorithms and their compliance with the automated brain perfusion ROI detection on T2-weighted MR images of a human head. The end part of this section describes the data used in the experiments. Next, the section Results and Discussion provides details on the setup of the experiments, then gives the experimental results and their discussion. Finally, section Conclusion completes the paper and references are at the end.

**Material and Methods**

Any algorithm that uses thresholding technique for extraction of image pixels with low-level intensity as background defines the binary mask $M(x, y)$ for the thresholded image as follows:

$$M(x,y) = \begin{cases} 1, & \text{if } I(x,y) > t; \\ 0, & \text{if } I(x,y) \leq t, \end{cases}$$

where $I(x,y)$ — image intensity at point with coordinates $(x,y)$, $t$ — threshold value.

In overall, thresholding algorithms can be classified as either global or local thresholding based on the rules of the threshold value detection. Global thresholding algorithms use a single threshold value for the entire image processing, whereas local thresholding algorithms divide the processed image into sets of pixels (sub-images) and for each of them a separate threshold value is used.

The analysis of effectiveness of thresholding techniques was done in accordance with segmentation results obtained from four different algorithms that are threshold-based and can be applied for automated brain perfusion ROI detection on T2-weighted MR images of a human head.

The first algorithm to be used for the analysis of effectiveness of thresholding techniques was implemented according to Otsu method [16]. Otsu method is a type of global thresholding technique that is very popular and widely used in medical image processing. The algorithm automatically searches for clustering-based image threshold that minimizes the intra-class variance. The reason for such search is that variance is used as a measure of image region homogeneity (i.e., image regions with higher homogeneity have lower variance). In order to find out the threshold value that minimizes the intra-class variance, the algorithm considers all possible values of image intensities as threshold candidates and calculates the intra-class variance for each of the two classes under consideration: the class of image pixels below and above considered threshold. Intra-class variance is calculated as follows:

$$\sigma_w^2(t) = \omega_1(t) \cdot \sigma_1^2(t) + \omega_2(t) \cdot \sigma_2^2(t),$$

where $\omega_1(t)$ and $\omega_2(t)$ — the probabilities of two classes separated by a threshold $t$; $\sigma_1^2(t)$ and $\sigma_2^2(t)$ — the variances of these two class. In order to decrease computation costs, the algorithm maximizes inter-class variance that is the same as minimizing the intra-class variance. The inter-class variance is obtained by extracting intra-class variance from the total variance and can be calculated as follows:

$$\sigma_b^2(t) = \sigma^2 - \sigma_w^2(t) = \omega_1(t) \cdot \omega_2(t) \cdot (\mu_1(t) - \mu_2(t))^2,$$

where $\omega_1(t)$ and $\omega_2(t)$ — the class probabilities are calculated from the $L$ bins of the image histogram as follows:

$$\omega_1(t) = \sum_{i=1}^{t} p(i)$$

and

$$\omega_2(t) = \sum_{i=t+1}^{L} p(i),$$

while the class means are calculated as follows:

$$\mu_1(t) = \frac{1}{\omega_1(t)} \cdot \sum_{i=1}^{t} p(i)x(i)$$

and

$$\mu_2(t) = \frac{1}{\omega_2(t)} = \sum_{i=t+1}^{L} p(i)x(i).$$

The second threshold-based algorithm to be used for the analysis was implemented according to Niblack method [17]. This method is a type of local thresholding technique, in which the threshold values are spatially varied and are calculated based on the local characteristics of the processed image. The algorithm searches for a local threshold value $t$ for pixel with coordinates $(x,y)$ within a window of size $w \times w$ as follows:

$$t(x,y) = \mu_w(x,y) + k \cdot \sigma_w(x,y),$$

where $\mu_w(x,y)$ and $\sigma_w(x,y)$ — the values of local mean and standard deviation of intensity values for all the pixels inside the search window respectively; $k$ — the bias that controls the level of adaptation varying the threshold value. Histogram equalization was made as a preprocessing step to improve Niblack method to be more effective in MR images thresholding [18].

The third algorithm to be used for the analysis is also threshold-based and uses approximate anatomical brain location (AABL) as image region for threshold value calculation [19]. The image processing by applying threshold value detected in AABL region is a type of global thresholding. Similar to the previous ones, the algorithm output is a binary mask of perfusion ROI that has zero values for air pixels and pixels that represent non-brain tissues. The algorithm searches for threshold value $t$ from AABL pixels as follows:

$$t = \mu_{AABL}(x,y) - \sigma_{AABL}(x,y),$$

where $\mu_{AABL}(x, y)$ and $\sigma_{AABL}(x, y)$ — the values of mean and standard deviation of intensity values for all the pixels inside the region of approximate anatomical brain location respectively. AABL region is obtained by cropping the processed image in global extrema places of the first derivative of the projection curves. The horizontal $P_H(x)$ and vertical $P_V(y)$ projection curves are the 1D functions of the standard deviation values obtained by projecting the image pixels onto horizontal or vertical axis as follows:

$$P_H(x) = \sqrt{\left(\sum_{j=1}^{M} I^2(x, j) - \frac{1}{N}\left(\sum_{j=1}^{M} I(x, j)\right)^2\right)/(N - 1)}$$

and

$$P_V(y) = \sqrt{\left(\sum_{j=1}^{N} I^2(j, y) - \frac{1}{M}\left(\sum_{j=1}^{N} I(j, y)\right)^2\right)/(M - 1)},$$

where $I(x, y)$ — image intensity at pixel with coordinates $(x, y)$; $N$ — the number of image columns; and $M$ — the number of image rows. Additionally, the algorithm uses hole filling [20] and binary region growing [21] steps to remove falsely detected regions and produce a region of only brain tissues.

The last one algorithm to be used for the analysis of effectiveness of thresholding techniques was implemented as a brute force search of the threshold value. For this purpose, thresholding was done with all possible values of image intensities as threshold candidates. The final threshold value was defined to present segmentation results with the highest similarity to the reference standard that was defined as manually marked ROI of the brain perfusion data by an experienced radiologist and confirmed by a second radiologist. Agreement with the reference standard was estimated with usage of Dice similarity index $DSI$, which value was calculated as follows:

$$DSI = \frac{2 \cdot TP}{2 \cdot TP + FP + FN},$$

where $FP$ — false positive pixels, which are defined as brain perfusion ROI pixels after thresholding applying, but they are not in the reference standard; $FN$ — false negative pixels, which are defined as pixels not of the brain perfusion ROI after thresholding applying, but they are in the reference standard; $TP$ — true positive pixels, which are defined as brain perfusion ROI pixels after thresholding applying, and they are in the reference standard.

The analysis of effectiveness of threshold-based techniques in brain perfusion ROI detection was performed on DSC perfusion MR images of a human head with abnormal brain anatomy from 12 patients with cerebrovascular disease.

The results shown here are in whole based upon data generated by the TCGA Research Network: http://cancergenome.nih.gov/.

All analyzed datasets were divided in two batches according to the scan parameters. The first batch (here, cases from 1 to 6) scan parameters were: repetition time = 1900 ms, echo time = 40 ms, flip angle = 90°, field of view = 23×23 cm, image size = = 128×128 pixels, voxel resolution = 1.875×11.875× ×15 mm³. The second batch (here, cases from 7 to 12) scan parameters were repetition time = 1550 ms, echo time = 40 ms, flip angle = 90°, field of view = = 23×123 cm, image size = 128×1128 pixels, voxel resolution = 1.875×11.875×16 mm³. Each analyzed MR dataset consisted of 5 slices with 95 dynamic images per slice. All images were collected in 12-bit DICOM (Digital Imaging and Communication in Medicine) format.

Image postprocessing software program was in-house developed to perform the analysis of effectiveness of different threshold-based techniques. It is written in C# and uses an open-source EvilDICOM (http://rexcardan.github.io/Evil-DICOM/) for loading medical images. The developed software program has no preprocessing, such as noise reduction, motion correction or intensity nonuniformity correction. Implementation of thresholding segmentation algorithms has no dependency on image resolution and is performed using the 4[th] time-point image, on which signal intensity is reached a steady state.

**Results and Discussion**

In the current study, the analysis was done using comparison of qualitative perfusion maps that were produced from segmented images and from the reference ones. Segmented images were obtained by applying four threshold-based algorithms to the original DSC head scans. Reference images were manually marked ROIs of the brain perfusion data by one experienced radiologist, and confirmed by another radiologist.

Pixels values for all images were changed to zero for background regions (air pixels and pixels that represent non-brain tissues) and were kept the same as original for foreground regions (brain perfusion ROIs).

The comparison was performed on the same DSC perfusion MR datasets of a human head with abnormal brain anatomy that were selected for the experiments. Each analyzed MR dataset consisted of 5 slices; all of them from each dataset were selected as the image set for the experiments. Consequently, the analysis of the effectiveness of threshold-based techniques in brain perfusion ROI detection was performed on 60 T2-weighted MR images with abnormal head anatomy.

Cerebral blood flow (CBF), cerebral blood volume (CBV), mean transit time (MTT), and time to maximum of residue function ($T_{max}$) maps were calculated by using a deconvolution algorithm. Arterial input function was determined by using the simplest pointing to the artery signals in the brain cross-section that was performed by one experienced radiologist and confirmed by another radiologist.

The analysis was done under the considered machining of perfusion ROIs with regions where analyzed images had non-zero pixel values.

In order to evaluate the effectiveness of threshold-based techniques in brain perfusion ROI detection, the Pearson correlation coefficient r was calculated to determine correlation between CBF, CBV, MTT, and $T_{max}$ perfusion maps from segmented images and from the reference. A difference with a p-value of less than 0.01 was considered significant for all experiments.

The slope and intercept of linear regression were also determined to evaluate the relationship between the perfusion maps from segmented images and from the reference.

Results of the analysis of effectiveness of threshold-based techniques in brain perfusion ROI detection on T2-weighted MR images with abnormal head anatomy are shown in two tables: Pearson correlation results are shown in Table 1, linear regression — in Table 2.

*Table* **1.** Correlation between the perfusion maps that were produced from the segmented images and from the reference ones. Data presented are Pearson correlation coefficients, $p$-value $<$ 0.01 for significance of correlation

| Map | Thresholding method | | | |
|---|---|---|---|---|
| | Otsu | Niblack | AABL | Brute force |
| CBF | 0.8128 | 0.7229 | 0.8311 | 0.8400 |
| CBV | 0.8026 | 0.7123 | 0.8105 | 0.8258 |
| MTT | 0.8441 | 0.7738 | 0.8469 | 0.8518 |
| $T_{max}$ | 0.3441 | 0.3494 | 0.3452 | 0.3499 |

As can be observed from the obtained results, brute force thresholding produced perfusion maps with the highest correlation to the reference in all cases. In most cases, the Otsu and AABL thresholding results were close to the brute force one. However, perfusion maps produced from the images segmented with AABL thresholding had a little bit better correlation with reference. Niblack thresholding showed the worst results in all cases. It should be mentioned that $T_{max}$ map had the highest subjection from segmentation among other perfusion maps. Despite the relatively high correlation results for CBF, CBV, and MTT maps, correlation with reference in case of $T_{max}$ was unacceptably poor for all thresholding methods.

The ideal condition would be to have slope and intercept of linear regression with reference as 1 and 0, respectively. However, the results of the regression analysis indicated that 95 % confidence intervals of

*Table* **2.** Linear regression analysis of the perfusion maps that were produced from the segmented images and from the reference ones: $y = ax + b$, $x$ = reference value, $y$ = measured value, $a$ = slope, $b$ = intercept. Data presented are 95 % confidence intervals for regression coefficients

| | Clinical case | Thresholding method | | | |
|---|---|---|---|---|---|
| | | Otsu | Niblack | AABL | Brute force |
| Slope | CBF | 0.7784 ± 0.0733 | 1.3691 ± 0.1466 | 0.7801 ± 0.0712 | 0.7844 ± 0.0621 |
| | CBV | 0.7262 ± 0.1474 | 1.0327 ± 0.1735 | 0.7298 ± 0.1408 | 0.7447 ± 0.1389 |
| | MTT | 0.8607 ± 0.03 | 0.7363 ± 0.0287 | 0.8625 ± 0.0299 | 0.8660 ± 0.0308 |
| | $T_{max}$ | 0.6426 ± 0.0988 | 0.6208 ± 0.1058 | 0.6431 ± 0.1044 | 0.6467 ± 0.0946 |
| Intercept | CBF | −0.6791 ± 1.5882 | 4.4029 ± 2.9543 | −0.6696 ± 1.5816 | −0.6414 ± 0.7827 |
| | CBV | 0.3077 ± 0.1747 | 0.4931 ± 0.1911 | 0.2924 ± 0.1745 | 0.2843 ± 0.1714 |
| | MTT | 0.1404 ± 0.0346 | 0.1865 ± 0.0396 | 0.1382 ± 0.0354 | 0.1278 ± 0.031 |
| | $T_{max}$ | 1.0507 ± 0.2457 | 1.4748 ± 0.3732 | 1.0488 ± 0.256 | 1.0435 ± 0.2735 |

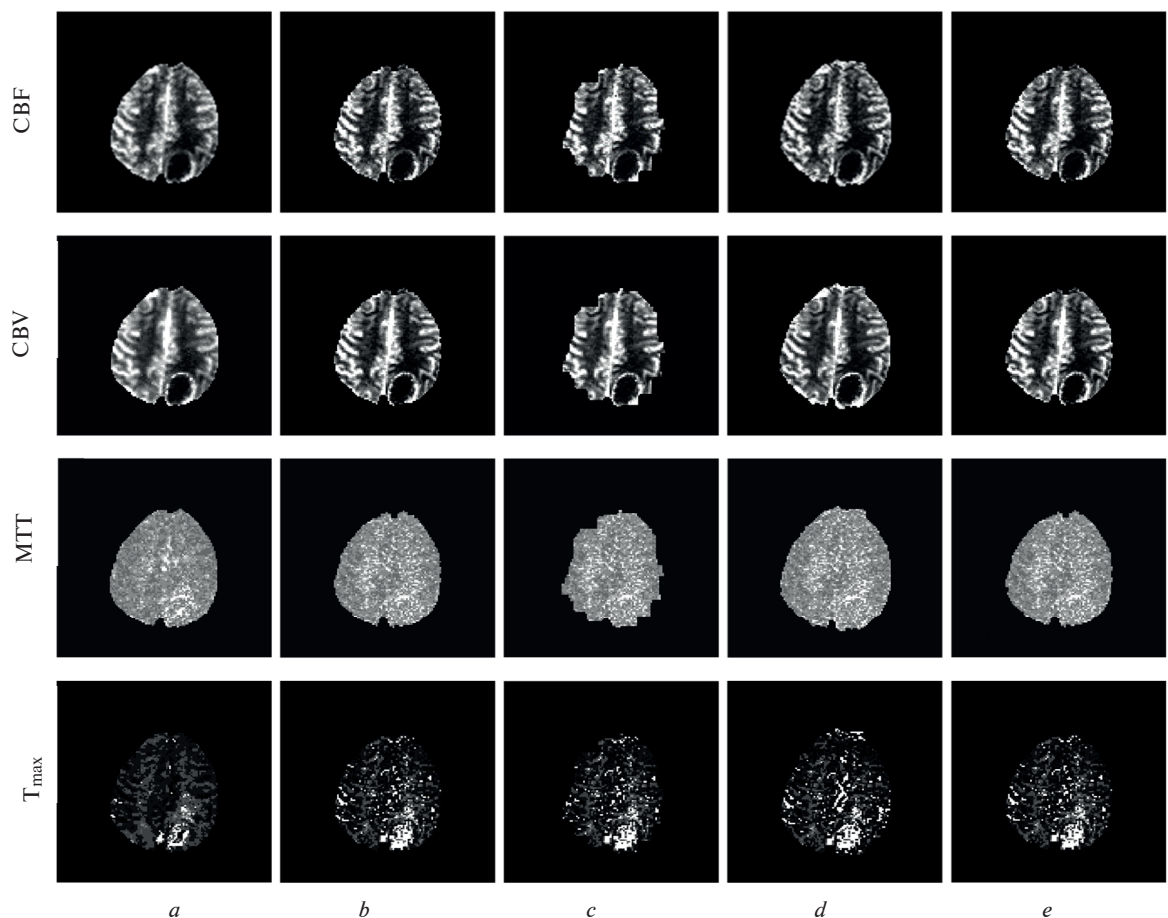

Perfusion maps, examples generated by using the same deconvolution techniques for different brain perfusion ROIs in one subject (all maps are shown with the same window/level settings): *a* — manually marked ROI (reference); *b* — Otsu thresholding; *c* — Niblack thresholding; *d* — thresholding in AABL region; *e* — brute force thresholding slope and intercept values were away from 1 and 0 in most cases. Consequently, perfusion maps produced from thresholded images were subject to scale and offset errors.

An example of perfusion maps produced from reference and thresholded images for representative case with abnormal brain anatomy is shown in the Figure.

**Conclusions**

The effectiveness of thresholding techniques in brain perfusion ROI detection on T2-weighted MR images of a human head with abnormal brain anatomy was analyzed in the current study. The analysis was done with the use of four different threshold-based algorithms that are widely applied for medical image processing (implementations of Otsu method as global thresholding and Niblack method as local thresholding), specifically developed for brain segmentation on T2-weighted MR images (thresholding in approximate anatomical brain location), and brute force thresholding to present segmentation results with the highest similarity to the reference standard. The analysis was performed on 60 T2-weighted MR images obtained from 12 patients with cerebrovascular disease. Quality of detected brain perfusion ROIs was considered with perfusion maps nature. Therefore, thresholded images, as well as reference ones, were used to produce CBF, CBV, MTT, and $T_{max}$ qualitative perfusion maps by deconvolution algorithm.

Although Pearson correlation analysis showed acceptable positive relationship between CBF, CBV, and MTT perfusion maps from thresholded images and from the reference in all conducted experiments ($r$ was ranged from 0.7123 to 0.8518, $p < 0.01$), correlation was weak in case of experiments with $T_{max}$ map ($r < 0.35$, $p < 0.01$). Linear regression analysis indicated that perfusion maps produced from thresholded images were subject to scale and offset errors at 95% level of confidence.

In conclusion, current study results have demonstrated that widely used thresholding methods are an ineffective way of managing pixels involved in brain perfusion ROI. Any manual turning of the threshold value to provide more accurate brain perfusion ROI detection can easily lead to degradation of perfusion maps quality. As can be seen from the performed analysis even insignificant difference in detected perfusion ROIs can cause a considerable drop in perfusion analysis results. Furthermore, it should be pointed out that perfusion ROI detection, similar to utilized LUT scheme and displayed values range of perfusion maps, has to be standardized quality control in perfusion analysis.

Є.М. Алхімова, С.В. Слюсар

АНАЛІЗ ЕФЕКТИВНОСТІ ПОРОГОВОЇ ФІЛЬТРАЦІЇ В ЗАДАЧІ ВИЗНАЧЕННЯ ЗОНИ ПЕРФУЗІЇ НА Т2-ЗВАЖЕНИХ МАГНІТНО-РЕЗОНАНСНИХ ПЕРФУЗІЙНИХ ЗОБРАЖЕННЯХ МОЗКУ З АНОМАЛЬНОЮ АНАТОМІЄЮ

**Проблематика.** Визначення області перфузії головного мозку є попереднім етапом перфузійного аналізу, який призначений для виключення пікселів, що не характеризують мозок, із зображень динамічно-сприйнятливої контрастної магнітно-резонансної (МР) томографії. Точність цього етапу вважається ключовим фактором у наданні правильних результатів перфузійного аналізу. Незважаючи на велику кількість алгоритмів сегментації мозку, не існує таких, які б точно і надійно працювали на Т2-зважених МР-зображеннях мозку людини з аномальною анатомією. Отже, порогова фільтрація, як і раніше, залишається тим способом, що широко використовується в сучасному програмному забезпеченні з перфузійного аналізу для визначення пікселів, які характеризують область перфузії головного мозку.

**Мета дослідження.** Аналіз ефективності методів порогової фільтрації щодо визначення області перфузії головного мозку на Т2-зважених МР-зображеннях мозку людини з аномальною анатомією.

**Методика реалізації.** Розглянуто чотири алгоритми пошуку порога: глобальний пошук за методом Оцу, локальний пошук за методом Ніблака, пошук у ділянці анатомічного розташування мозку і пошук за методом перебору. Результатом роботи всіх алгоритмів було зображення із заміною нульовими значеннями пікселів фону (пікселів повітря і пікселів, що не характеризують мозок) і з оригінальними значеннями пікселів з області перфузії головного мозку. Аналіз проводили, порівнюючи перфузійні карти, що були отримані із зображень після застосування порогової фільтрації та з еталонних зображень (мануальне визначення ділянки мозку досвідченими рентгенологами). Для порівняння були використані одні й ті самі зображення динамічно-сприйнятливої контрастної МР-томографії головного мозку 12 пацієнтів із цереброваскулярними захворюваннями.

**Результати дослідження.** Кореляційний аналіз Пірсона показав сильний позитивний ($r$ був від 0,7123 до 0,8518, $p < 0,01$) і слабкий позитивний ($r < 0,35$, $p < 0,01$) взаємозв'язок у проведених експериментах із CBF, CBV, MTT і $T_{max}$ перфузійними картами відповідно. Лінійний регресійний аналіз показав, що перфузійні карти, які були отримані із зображень після застосування порогової фільтрації, схильні до помилок масштабу і зсуву в усіх проведених експериментах з урахуванням 95 %-ного довірчого інтервалу.

**Висновки.** Результати експериментів показали, що поширене використання порогової фільтрації є неефективним способом визначення пікселів, які характеризують область перфузії головного мозку. Використання порогової фільтрації як інструменту з проведення сегментації мозку може призводити до неправильного визначення області перфузії, і, як наслідок, перфузійні карти будуть схильні до наявності артефактів і призведуть до помилково високої або помилково низької оцінки параметрів перфузії.

**Ключові слова:** перфузійна динамічно-сприйнятлива контрастна магнітно-резонансна томографія; зрізи з аномальною анатомією мозку; зона уваги; сегментація; порогова фільтрація.

С.Н. Алхимова, С.В. Слюсарь

АНАЛИЗ ЭФФЕКТИВНОСТИ ПОРОГОВОЙ ФИЛЬТРАЦИИ В ЗАДАЧЕ ОПРЕДЕЛЕНИЯ ОБЛАСТИ ПЕРФУЗИИ НА Т2-ВЗВЕШЕННЫХ МАГНИТНО-РЕЗОНАНСНЫХ ПЕРФУЗИОННЫХ ИЗОБРАЖЕНИЯХ МОЗГА С АНОМАЛЬНОЙ АНАТОМИЕЙ

**Проблематика.** Определение области перфузии головного мозга является предварительным этапом перфузионного анализа, который предназначен для исключения не относящихся к мозгу пикселей из изображений динамично-восприимчивой контрастной магнитно-резонансной (МР) томографии. Точность этого этапа считается ключевым фактором в предоставлении правильных результатов перфузионного анализа. Несмотря на большое разнообразие алгоритмов сегментации мозга, не существует таких, которые бы точно и надежно работали на Т2-взвешенных МР-изображениях мозга человека с аномальной анатомией. Таким образом, пороговая фильтрация по-прежнему остается широко используемым способом определения пикселей, которые характеризуют область перфузии головного мозга, в современном программном обеспечении для проведения перфузионного анализа.

**Цель исследования.** Анализ эффективности методов пороговой фильтрации в определении области перфузии головного мозга на Т2-взвешенных МР-изображениях мозга человека с аномальной анатомией.

**Методика реализации.** Рассмотрены четыре алгоритма поиска порога: глобальный поиск методом Оцу, локальный поиск методом Ниблака, поиск в области анатомического расположения мозга и поиск методом перебора. Результатом всех алгоритмов являлось изображение с измененными на ноль значениями пикселей для фона (пиксели воздуха и пиксели, которые представляют ткани, не являющиеся мозгом) и оригинальными значениями для пикселей области перфузии головного мозга. Анализ проводился на основании сравнения перфузионных карт, полученных из отсегментированных пороговой фильтрацией изображений и из эталонных (мануальное определение области мозга опытными рентгенологами). Для сравнения были использованы одни и те же изображения динамично-восприимчивой контрастной МР-томографии головного мозга 12 пациентов с цереброваскулярными заболеваниями.

**Результаты исследования.** Корреляционный анализ Пирсона показал сильную положительную ($r$ был от 0,7123 до 0,8518, $p < 0,01$) и слабую положительную ($r < 0,35$, $p < 0,01$) взаимосвязь в случае проведенных экспериментов с CBF, CBV, MTT и $T_{max}$ перфузионными картами соответственно. Линейный регрессионный анализ показал, что перфузионные карты, которые были получены из отсегментированных пороговой фильтрацией изображений, подвержены ошибкам масштаба и смещения во всех проведенных экспериментах с учетом 95 %-ного доверительного интервала.

**Выводы.** Результаты экспериментов показали, что широко используемые методы пороговой фильтрации являются неэффективным способом определения пикселей, которые характеризуют область перфузии головного мозга. Использование пороговой фильтрации как инструмента для проведения сегментации мозга может приводить к неправильному определению области перфузии, и, как следствие, перфузионные карты будут подвержены наличию артефактов и приведут к ошибочно высокой или ошибочно низкой оценке параметров перфузии.

**Ключевые слова:** перфузионная динамично-восприимчивая контрастная магнитно-резонансная томография; срезы с аномальной анатомией мозга; зона интереса; сегментация; пороговая фильтрация.